\documentclass[pra,aps,10pt,twocolumn,showpacs,amsmath,amssymb]{revtex4-1}

\usepackage[english]{babel}
\usepackage{afterpage}
\usepackage{graphicx}
\usepackage{dcolumn}
\usepackage{bm}
\usepackage{color}
\usepackage{slashed}
\usepackage{latexsym}

\newcommand{\op}[1]{\hat{#1}} 
\newcommand{\opb}[1]{\hat{\bm{#1}}} 

\newcommand{\dd}{d}         
\newcommand{\ii}{i}  
\newcommand{\ee}{e}  

\newcommand{\bra}[1]{\langle{#1}\vert} 
\newcommand{\ket}[1]{\vert{#1}\rangle} 
\newcommand{\scal}[2]{\langle #1 | #2 \rangle} 

\begin{document}

\title{Generation of electron vortex states in ionization by intense and short laser pulses}
\author{F. Cajiao V\'elez}
\email[E-mail address:\;]{Felipe.Cajiao-Velez@fuw.edu.pl}
\author{K. Krajewska}
\email[E-mail address:\;]{Katarzyna.Krajewska@fuw.edu.pl}
\author{J. Z. Kami\'nski}
\affiliation{Institute of Theoretical Physics, Faculty of Physics, University of Warsaw, Pasteura 5,
02-093 Warsaw, Poland
}
\date{\today}

\begin{abstract}
The generation of electron vortex states in ionization by intense and short laser pulses is analyzed under the scope of the lowest-order Born approximation. 
For near infrared laser fields and nonrelativistic intensities of the order of $10^{16}$~W/cm$^2$, we show that one has to modify the nonrelativistic 
treatment of ionization by accounting for recoil and mass relativistic corrections. By using the corrected quasi-relativistic theory, the requirements for the observation 
of electron vortex states with non-negligible probability and large topological charge are determined. 
\end{abstract}

\maketitle

\section{Introduction}
\label{sec::intro}

Even though the optical vortices have been studied theoretically already in the 40's of the last century~\cite{Stratton1941}, 
their usefulness in modern physics and technology has been realized quite recently. Both theoretical and experimental advances in the field of optical and
matter vortex waves have been reviewed in recent articles devoted to light~\cite{Review3} and electrons~\cite{Review1,Review2}, and in the reviews 
on Bose-Einstein condensates~\cite{Fetter2009,SRHM2010} and quantum fluids of light~\cite{CC2013}.

In this paper, we consider the electron vortex states (EVS) generated in strong-field ionization. Actually, the EVS in quantum mechanics can 
arise in different scenarios. For instance, they manifest themselves in the quantum Hall, De~Haas-van~Alphen, and Shubnikov-De~Hass effects as 
collective properties of condensed-matter electrons in solids~\cite{Kittel1987,Abrikosov1988,Marder2010}. Related to this is the appearance
of impurity resonant states in two-dimensional quantum wells~\cite{KKvortex1,KKvortex2}. Such states, observed in crossed magnetic and electric fields, 
have the vortex-like structure. Positions of their singularities (i.e., points at which the electron wave function vanishes and
its phase is not uniquely defined~\cite{BCK}) can be controlled by external fields. Namely, for certain
field parameters, the vortex singularities can be aligned and the usually short-living resonances become the long-living 
ones~\cite{KKvortex1,KKvortex2}. Note that the propagation of EVS in magnetic fields has been also discussed in both the Aharonov-Bohm 
and Landau configurations~\cite{BSVN2012}, showing their distinctive phase properties. The creation of EVS in angle-resolved 
photoemission of electrons from solids and their relation to the Berry phase has been studied in Ref.~\cite{TN2015}. 
Moreover, the Stern-Gerlach-type measurement of electrons with large orbital angular momenta has been analyzed~\cite{HGM2017}. 
It is particularly important in light of the current paper that EVS can be generated in laser-assisted quantum processes, such as scattering~\cite{SSF2014} and 
ionization~\cite{DHMMMS2015,DGGSVB2016,DMMHMS2017,PKJEBW2017} in strong laser fields. Supplementary to these investigations is the analysis of electron 
recombination~\cite{MHSSF2014} and scattering~\cite{Ivanov2012,ISSF2016,KKSS2017} in the absence of the laser pulse, or the propagation of EVS 
in a strong laser wave~\cite{BG2005,Karlovets2012,HMASF2014,BBC2015}. In addition, free-electron vortex states have been studied recently in~\cite{BDN2011,BB2017,Barnett2017}.

The aim of this paper is to investigate ionization by intense and short laser pulses resulting in electron states of very large orbital angular momenta. 
For this purpose, we shall focus on the high-energy portion of the ionization spectrum. In order to neglect spin effects, we limit ourselves
to nonrelativistic laser pulse intensities of roughly $10^{16}$~W/cm$^2$. The reason being that, for high-energy ionization by near infrared laser fields, 
the spin corrections are marginal at these intensities. We find, however, that other corrections (such as the recoil and relativistic mass corrections) already play 
a role and have to be incorporated into the nonrelativistic theory. Our new quasi-relativistic treatment is an extension of~\cite{KKpress}
and, in the regime of parameters considered in the current paper, agrees very well with the fully relativistic approach. In order to select optimal conditions for the generation of EVS we shall discuss the 
concept of the ionization spiral, around which the probability distribution of high-energy ionization is peaked. We will show that, if the electron momenta 
of vortex states follow the spiral, the EVS of large orbital angular momenta are created with significant probabilities.

The paper is organized as follows. While in Sec.~\ref{sec:transprob} we define the transition probabilities for arbitrarily 
normalized states, in Sec.~\ref{sec::planewave} we apply this general scheme to the plane-wave states of well defined momenta. Some properties of EVS, 
together with the notation used in this paper, are discussed in Sec.~\ref{sec:twistfree}. Also, the transition probabilities and amplitudes involving EVS 
are discussed there. The generalization to the electron scattering vortex states for static and spherically symmetric potentials is elucidated in 
Sec.~\ref{sec:ScatteringTwist}. Ionization of a one-electron system is discussed in Sec.~\ref{sec:probdist}. In particular, in Sec.~\ref{sec:theory}, 
we derive the exact differential probability distribution of ionization to a vortex state. The lowest-order Born approximation is discussed in Sec.~\ref{sec:CSFA}, 
together with the importance of the recoil and mass corrections. In Secs.~\ref{model} and~\ref{sec:comparison}, we define the shape of the laser 
pulse and introduce two quasi-relativistic approximations. We show that, in the high-energy portion of the ionization spectrum (i.e., for kinetic energies of the order 
of 1~keV or larger), the relativistic mass corrections become significant for the considered Ti:Sapphire laser field. Sec.~\ref{sec:twist} is devoted to the 
creation of EVS. In order to generate such states efficiently in strong-field ionization, it is necessary to choose properly the parameters 
of the final electron momenta. Namely, they have to follow the ionization spiral which is discussed in Sec.~\ref{sec:ionspiral}. Probability distributions
of EVS as well as their properties are analyzed in Sec.~\ref{sec:OAM}. Finally, in Sec.~\ref{sec:Conclusions} we summarize our results and draw perspectives 
for further investigations.

Throughout the paper we keep $\hbar=1$. Unless otherwise stated, in our numerical 
analysis we use relativistic units (rel. units) such that $\hbar=m_{\rm e}=c=1$, where $m_{\rm e}$ is the electron rest mass.

\section{Transition probabilities}
\label{sec:transprob}

Let us start with the most general situation when the time-evolution of a system is described by a unitary operator $\op{S}$, $(\op{S}^{\dagger}\op{S}=\op{I})$ and, 
in the remote past, it is prepared in an initial state $\ket{\mathrm{in}}$ such that $\scal{\mathrm{in}}{\mathrm{in}}=N_{\mathrm{in}}<\infty$. We further assume that 
in the far future the Hilbert space of the system is spanned by a set of orthogonal states $\ket{\Lambda}$,
\begin{equation}
\scal{\Lambda'}{\Lambda}=N_{\Lambda}\delta_{\Lambda'\Lambda},
\label{tp1}
\end{equation}
that satisfy the completeness relation,
\begin{equation}
\sum_{\Lambda}\frac{1}{N_{\Lambda}}\ket{\Lambda}\bra{\Lambda}=\op{I}.
\label{tp2}
\end{equation}
In general, $\Lambda$ is a multi-index labeling these states and it contains both continuous and discrete parameters. For the continuous parameters, 
the symbol $\delta_{\Lambda'\Lambda}$ in~\eqref{tp1} has to be replaced by the Dirac delta distribution, whereas the sum over $\Lambda$ in Eq.~\eqref{tp2} refers to integration.

The transition probability amplitude from the initial $\ket{\mathrm{in}}$ to the final state $\ket{\Lambda}$ is defined as the matrix element 
of the corresponding evolution operator $\op{S}$,
\begin{equation}
\mathcal{A}_{\mathrm{in}}(\Lambda)=\bra{\Lambda}\op{S}\ket{\mathrm{in}}.
\label{tp3}
\end{equation}
It follows from the unitarity of $\op{S}$ that these amplitudes satisfy the sum rule,
\begin{equation}
\sum_{\Lambda}\frac{1}{N_{\mathrm{in}}N_{\Lambda}}|\mathcal{A}_{\mathrm{in}}(\Lambda)|^2=1.
\label{tp4}
\end{equation}
Hence, the transition probabilities are equal to
\begin{equation}
\mathcal{P}_{\mathrm{in}}(\Lambda)=\frac{1}{N_{\mathrm{in}}N_{\Lambda}}|\mathcal{A}_{\mathrm{in}}(\Lambda)|^2.
\label{tp5}
\end{equation}
Of course, the Hilbert space of the system can be spanned by a different set of orthogonal and complete states $\ket{\Xi}$, labeled by a multi-index $\Xi$.
In this case, the corresponding transition probabilities are
\begin{equation}
\mathcal{P}_{\mathrm{in}}(\Xi)=\frac{1}{N_{\mathrm{in}}N_{\Xi}}|\mathcal{A}_{\mathrm{in}}(\Xi)|^2,
\label{tp6}
\end{equation}
with
\begin{equation}
\mathcal{A}_{\mathrm{in}}(\Xi)=\sum_{\Lambda}\frac{\scal{\Xi}{\Lambda}}{N_{\Lambda}}\mathcal{A}_{\mathrm{in}}(\Lambda).
\label{tp7}
\end{equation}
This defines how to transform the probability amplitudes when calculated in different bases. To illustrate this general approach, we consider now free-electron states.

\subsection{Free-electron plane-wave states}
\label{sec::planewave}

For a free electron, the plane-wave states $\ket{\bm{p}}$, where $\bm{p}$ is the electron momentum, are the most common choice of the states $\ket{\Lambda}$. 
Their wave function in position representation are 
\begin{equation}
\scal{\bm{x}}{\bm{p}}=\ee^{\ii\bm{p}\cdot\bm{x}}.
\label{tp8}
\end{equation}
Hence,
\begin{align}
\scal{\bm{p}'}{\bm{p}}=\int \dd^3x\, \scal{\bm{p}'}{\bm{x}}\scal{\bm{x}}{\bm{p}}&=(2\pi)^3\delta^{(3)}(\bm{p}-\bm{p}'), \label{tp9}\\
\frac{1}{(2\pi)^3}\int\dd^3p\scal{\bm x}{\bm p}\scal{\bm p}{{\bm x}'}&=\delta^{(3)}(\bm{x}-\bm{x}'), \label{tp9new}
\end{align}
and the transition probability distribution,
\begin{equation}
\mathcal{P}_{\mathrm{in}}(\bm{p})=\frac{1}{(2\pi)^3N_{\mathrm{in}}}|\mathcal{A}_{\mathrm{in}}(\bm{p})|^2=\frac{1}{(2\pi)^3N_{\mathrm{in}}}|\bra{\bm{p}}\op{S}\ket{\mathrm{in}}|^2,
\label{tp10}
\end{equation}
satisfies the completeness relation,
\begin{equation}
\int\dd^3p\, \mathcal{P}_{\mathrm{in}}(\bm{p})=1.
\label{tp11}
\end{equation}

\subsection{Free-electron vortex states}
\label{sec:twistfree}

Now, we choose a different basis of free-electron states. In order to define them, we choose first an arbitrary unit vector in space $\bm{n}_{\|}$, 
which is uniquely determined by the polar and azimuthal angles $\theta_{\mathrm{T}}$ and $\varphi_{\mathrm{T}}$, respectively. This vector together with
two other vectors, $\bm{n}_{\bot,1}$ and $\bm{n}_{\bot,2}$, 
\begin{align}
\bm{n}_{\bot,1}=&\begin{pmatrix}
\cos\theta_{\mathrm{T}}\cos\varphi_{\mathrm{T}} \cr \cos\theta_{\mathrm{T}}\sin\varphi_{\mathrm{T}} \cr -\sin\theta_{\mathrm{T}}
\end{pmatrix}, \,
\bm{n}_{\bot,2}=\begin{pmatrix}
-\sin\varphi_{\mathrm{T}} \cr \cos\varphi_{\mathrm{T}} \cr 0
\end{pmatrix}, \nonumber \\
 \bm{n}_{\|}=&\begin{pmatrix}
\sin\theta_{\mathrm{T}}\cos\varphi_{\mathrm{T}} \cr \sin\theta_{\mathrm{T}}\sin\varphi_{\mathrm{T}} \cr \cos\theta_{\mathrm{T}}
\end{pmatrix},
\label{twisttriad}
\end{align}
constitute a triad of right-handed orthogonal unit vectors~\cite{KK2014a,KK2014b}.
The new states are defined as free-electron states which are eigenvectors of $\op{L}_{\|}=\bm{n}_{\|}\cdot \opb{L}$, where $\opb{L}=\opb{x}\times\opb{p}$ is the 
orbital angular momentum operator. We will refer to them as {\it free-electron vortex states}.

The triad of vectors~\eqref{twisttriad} defines a cylindrical coordinate system, in which a position vector $\bm{x}$ can be decomposed such that
\begin{equation}
\bm{x}=x_{\|}\bm{n}_{\|}+x_{\bot}(\bm{n}_{\bot,1}\cos\varphi_x+\bm{n}_{\bot,2}\sin\varphi_x),
\label{tw1}
\end{equation}
and similarly for a momentum vector ${\bm p}$. One can show that the free-electron vortex states $\ket{p_{\|},p_{\bot},m}$ (also called the twisted or Bessel states)
in position representation have the form
\begin{equation}
\scal{\bm{x}}{p_{\|},p_{\bot},m}=\ii^m \ee^{\ii p_{\|}x_{\|}}J_m(p_{\bot}x_{\bot})\ee^{\ii m\varphi_x},
\label{tw3}
\end{equation}
where the parallel and perpendicular components of the electron momentum are
\begin{equation}
p_{\|}=\bm{p}\cdot\bm{n}_{\|} \quad \textrm{and} \quad p_{\bot}=\sqrt{\bm{p}^2-p_{\|}^2},
\label{tw2}
\end{equation}
and the integer $m$ is called the topological charge. The free-electron wave functions~\eqref{tw3} fulfill the orthogonality condition~\eqref{tp1},
\begin{equation}
\scal{p'_{\|},p'_{\bot},m'}{p_{\|},p_{\bot},m}=\frac{(2\pi)^2}{p_{\bot}}\delta(p_{\|}-p'_{\|})\delta(p_{\bot}-p'_{\bot})\delta_{mm'},
\label{tw4}
\end{equation}
which follows from the property of the Bessel functions (see, e.g.,~\cite{TS1972,AG1993,Sneddon}),
\begin{equation}
\int_0^{\infty} x_{\bot}\dd x_{\bot}\, J_m(p'_{\bot}x_{\bot})J_m(p_{\bot}x_{\bot})=\frac{1}{p_{\bot}}\delta(p_{\bot}-p'_{\bot}).
\label{tw6}
\end{equation}
Hence, the following completeness relation~\eqref{tp2} for the wave functions~\eqref{tw3} holds
\begin{align}
\frac{1}{(2\pi)^2}\sum_{m=-\infty}^{\infty}\int_{-\infty}^{\infty}\dd p_{\|}&\int_0^{\infty}p_{\bot}\dd p_{\bot}
\scal{\bm{x}'}{p_{\|},p_{\bot},m}
\nonumber \\
\times &\scal{p_{\|},p_{\bot},m}{\bm{x}}=\delta^{(3)}(\bm{x}-\bm{x}').
\label{tw5}
\end{align}

\begin{figure}
\includegraphics[width=6cm]{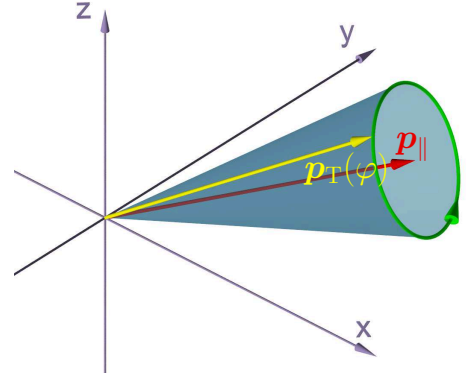}
\caption{The twisted momentum $\bm{p}_{\rm T}(\varphi)$ circulating on the lateral surface of the cone with apex at the origin of coordinates, the opening angle 
$2\beta_{\mathrm{T}}$, and its side length $p_{\mathrm{T}}$. The symmetry axis is defined by the polar and azimuthal angles $\theta_{\mathrm{T}}$ and $\varphi_{\mathrm{T}}$, 
respectively. Here, $\bm{p}_{\rm T}(\varphi)$ is parametrized by the angle $0\leqslant \varphi\leqslant 2\pi$ [see, Eq.~\eqref{twistmom1}] for $\zeta_H=1$. While the momentum 
$\bm{p}_{\|}$, parallel to the symmetry axis and fixed at the cone's apex, is independent of $\varphi$ and has the length $p_{\|}=p_{\mathrm{T}}\cos\beta_{\mathrm{T}}$, 
the perpendicular component $\bm{p}_{\bot}(\varphi)$ rotates on the cone's circular base of radius $p_{\bot}=p_{\mathrm{T}}\sin\beta_{\mathrm{T}}$.
}
\label{twistmomentum00}
\end{figure}

Now, our aim is to determine the probability amplitude of a transition to a free-electron vortex state~\eqref{tw3} knowing the respective probability amplitudes
to the plane-wave states (see, Sec.~\ref{sec::planewave}). Since the EVS wave functions~\eqref{tw3} depend on the choice of the coordinate system, which is defined by
the angles $\theta_{\rm T}$ and $\varphi_{\rm T}$, we will attach the same subscript to the momentum labeling the plane waves, $\ket{{\bm p}_{\rm T}(\varphi)}$.
This is to emphasize that these states are determined in the cylindrical coordinates~\eqref{twisttriad}. In this case, 
the electron momentum $\bm{p}_{\rm T}(\varphi)$ can be parametrized by the angle $\varphi\in [0,2\pi]$ (see, Fig.~\ref{twistmomentum00}),
\begin{align}
\bm{p}_{\rm T}(\varphi)=&\bm{p}_{\|}+\bm{p}_{\bot}(\varphi)=p_{\mathrm{T}}\cos\beta_{\mathrm{T}}\bm{n}_{\|} \nonumber \\
 &+p_{\mathrm{T}}\sin\beta_{\mathrm{T}}(\bm{n}_{\bot,1}\cos\varphi+\zeta_H \bm{n}_{\bot,2}\sin\varphi).
\label{twistmom1}
\end{align}
Here, we understand that the momentum $\bm{p}_{\|}$ is parallel to the axis ${\bm n}_{\|}$ and it has the origin at the cone's apex; therefore, it
is independent of $\varphi$. The perpendicular component $\bm{p}_{\bot}(\varphi)$, on the other hand, rotates on the cone's circular base of radius 
$p_{\mathrm{T}}\sin\beta_{\mathrm{T}}$. The direction of rotation is controlled by the sign of $\zeta_H=\pm$, which determines the helicity of the vortex state. 
Without loosing generality, we assume that $\zeta_H=1$. Also, we will call the momenta~\eqref{twistmom1} 
the \textit{family of twisted momenta} and the parameter $\varphi$ the \textit{twist angle}.

Now, by applying \eqref{tw1} and the generating function for the Bessel functions,
\begin{equation}
\ee^{\ii x\cos\varpi}=\sum_{m=-\infty}^{\infty}\ii^m J_m(x)\ee^{\ii m\varpi},
\label{tw8}
\end{equation}
we find out that the state $|\bm{p}_{\rm T}(\varphi)\rangle$, in position representation, can be expanded as
\begin{equation}
\scal{\bm{x}}{\bm{p}_{\rm T}(\varphi)}=\ee^{\ii \bm{x}\cdot\bm{p}_{\rm T}(\varphi)}=\sum_{m=-\infty}^{\infty}\ee^{-\ii m\varphi}\scal{\bm{x}}{p_{\|},p_{\bot},m},
\label{tw9}
\end{equation}
where $p_{\|}=p_{\mathrm{T}}\cos\beta_{\mathrm{T}}$ and $p_{\bot}=p_{\mathrm{T}}\sin\beta_{\mathrm{T}}$, in accordance with the definition~\eqref{twistmom1}. 
It follows from Eq.~\eqref{tw9} that
\begin{equation}
\ket{p_{\|},p_{\bot},m}=\frac{1}{2\pi}\int_0^{2\pi}\dd\varphi\, \ee^{\ii m\varphi}\ket{\bm{p}_{\rm T}(\varphi)}.
\label{tw10}
\end{equation}
Hence, the transition probability amplitude to the vortex state, $\mathcal{A}_{\mathrm{in}}(p_{\|},p_{\bot},m)$, can be expressed in terms of the transition 
probability amplitudes to the plane-wave states, $\mathcal{A}_{\mathrm{in}}(\bm{p}_{\rm T}(\varphi))$, such that
\begin{equation}
\mathcal{A}_{\mathrm{in}}(p_{\|},p_{\bot},m)=\frac{1}{2\pi}\int_0^{2\pi}\dd\varphi\, \ee^{-\ii m\varphi}\mathcal{A}_{\mathrm{in}}(\bm{p}_{\rm T}(\varphi)).
\label{tw11}
\end{equation}
For completeness, we also write that
\begin{equation}
\mathcal{A}_{\mathrm{in}}(\bm{p}_{\rm T}(\varphi))=\sum_{m=-\infty}^{\infty}\ee^{\ii m\varphi}\mathcal{A}_{\mathrm{in}}(p_{\|},p_{\bot},m) ,
\label{tw11a}
\end{equation}
which follows directly from Eq.~\eqref{tw9}.

Finally, according to the general formula~\eqref{tp5}, we arrive at the transition probability distribution to the free-electron vortex state $\ket{p_{\|},p_{\bot},m}$,
\begin{equation}
\frac{\dd^2\mathcal{P}_m}{\dd p_{\|}\dd p_{\bot}}\equiv\mathcal{P}_m(p_{\|},p_{\bot})=\frac{p_{\bot}}{(2\pi)^2N_{\mathrm{in}}}|\mathcal{A}_{\mathrm{in}}(p_{\|},p_{\bot},m)|^2,
\label{tw12}
\end{equation}
where $\mathcal{A}_{\mathrm{in}}(p_{\|},p_{\bot},m)$ can be obtained from~\eqref{tw11}.

In order to describe ionization, which is the main topic of this paper, one has to calculate the transition to the final scattering state.
For this reason, we will demonstrate next that the same formulation as presented here for the free-electron states [cf., Eq.~\eqref{tw10}]
can be carried on with the scattering states of the electron.

\subsection{Scattering vortex states}
\label{sec:ScatteringTwist}

Consider the scattering states of an electron interacting with a static and spherically symmetric atomic potential. There are two types of such states: 
the ones with outgoing spherical waves, $\psi^{(+)}_{\bm{p}}(\bm{x})$, and the ones with incoming spherical waves, $\psi^{(-)}_{\bm{p}}(\bm{x})$~\cite{RodbergThaler1967}; 
both labeled by the asymptotic electron momentum ${\bm p}$. These two wave functions are not independent, since
\begin{equation}
\bigl[\psi^{(-)}_{\bm{p}}(\bm{x})\bigr]^*=\psi^{(+)}_{-\bm{p}}(\bm{x}).
\label{tw14}
\end{equation}
Similar to~\cite{Taylor1972}, if considered in the abstract Hilbert space, we shall denote these states as $\ket{\bm{p};+}$ and $\ket{\bm{p};-}$, respectively. 
The question is: How to construct the corresponding scattering vortex states having known $\ket{{\bm p},\pm}$?
Based on Eq.~\eqref{tw14}, we understand that it is sufficient to define the scattering vortex state for either $\ket{\bm{p};+}$ or $\ket{\bm{p};-}$. We shall 
do this for the latter. The reason being that it is the scattering state with the incoming spherical waves that has to be accounted for in 
the transition probability amplitude of ionization. On the other hand, when analyzing recombination one should use $\ket{\bm{p};+}$ instead~\cite{JKE2000}.

For a spherically symmetric and static potential the time-independent Schr\"odinger equation is rotationally invariant. 
Since the boundary conditions imposed on the scattering states depend only on scalars with respect to rotations (i.e., $\bm{x}^2$, $\bm{p}^2$, and $\bm{p}\cdot\bm{x}$), 
the exact solution to the Schr\"odinger equation also depends only on these combinations. This property is used, for instance, in the partial wave analysis 
of scattering by a spherically symmetric potentials~\cite{BCK,Taylor1972}. The exact solution of scattering problem for the Coulomb potential can serve as 
an example of this general property. 

Having this in mind, we write the scattering state with incoming spherical waves, in position representation, as
\begin{equation}
\scal{\bm{x}}{\bm{p};-}=\psi^{(-)}_{\bm{p}}(\bm{x})=f_{\psi}^{(-)}(\bm{x}^2,\bm{p}^2,\bm{p}\cdot\bm{x}),
\label{tw15}
\end{equation}
where $f_{\psi}^{(-)}$ is \textit{a priori} unknown function of its arguments. In our case, the momentum in~\eqref{tw15} is the twisted
momentum ${\bm p}_{\rm T}(\varphi)$ [Eq.~\eqref{twistmom1}]. Since $\bm{p}^2_{\rm T}(\varphi)=p_{\|}^2+p_{\bot}^2$ and
\begin{equation}
\bm{p}_{\rm T}(\varphi)\cdot\bm{x}=p_{\|}x_{\|}+p_{\bot}x_{\bot}\cos(\varphi_x-\varphi),
\label{tw16}
\end{equation}
the wave function~\eqref{tw15} can be Fourier decomposed,
\begin{equation}
\scal{\bm{x}}{\bm{p}_{\rm T}(\varphi);-}=\sum_{m=-\infty}^\infty\ee^{-\ii m\varphi}\scal{\bm{x}}{p_{\|},p_{\bot},m;-}.
\label{tw16a}
\end{equation}
One can show that
\begin{equation}
\scal{\bm{x}}{p_{\|},p_{\bot},m;-}=\ee^{\ii m\varphi_x}f_{\psi,m}^{(-)}(x_{\|},x_{\bot};p_{\|},p_{\bot}),
\label{tw18}
\end{equation}
with
\begin{align}
f_{\psi,m}^{(-)} &(x_{\|},x_{\bot};p_{\|},p_{\bot})=\frac{1}{2\pi} \int_0^{2\pi}\dd\varpi \, \ee^{-\ii m\varpi} \\
&\times f_{\psi}^{(-)}\bigl(x_{\|}^2+x_{\bot}^2,p_{\|}^2+p_{\bot}^2,p_{\|}x_{\|}+p_{\bot}x_{\bot}\cos\varpi \bigr),  \nonumber
\label{tw19}
\end{align}
is an eigenfunction of the operator $\op{L}_{\|}=\bm{n}_{\|}\cdot\opb{L}$ with the eigenvalue $m$; hence, it defines the scattering vortex wave function with the incoming
spherical waves. As it follows from Eq.~\eqref{tw16a}, the scattering vortex state is
\begin{equation}
\ket{p_{\|},p_{\bot},m;-}=\frac{1}{2\pi}\int_0^{2\pi}\dd\varphi\, \ee^{\ii m\varphi}\ket{\bm{p}_{\rm T}(\varphi);-},
\label{tw17}
\end{equation}
which is an analogue of Eq.~\eqref{tw10}. As a consequence, for spherically symmetric potentials, the expressions for the amplitudes and probability distributions 
[Eqs.~\eqref{tw11} and \eqref{tw12}, respectively] remain unchanged if the plane-wave state $\ket{\bm{p}}$ is replaced by the scattering one $\ket{\bm{p};-}$.
Note also that, if the final energy of the electron is sufficiently large, the Born approximation can be applied. In its lowest order, this is equivalent to approximate 
the final scattering state by a plane wave. Hence, in the zeroth-order Born approximation, the function $f_{\psi}^{(-)}$ defined in~\eqref{tw15} becomes the plane 
wave $\ee^{\ii\bm{p}\cdot\bm{x}}$ and we recover the Bessel states discussed above.

\section{Ionization distributions}
\label{sec:probdist}

After these general remarks, we present now the theoretical treatment of strong-field ionization leading to generation of EVS. 

\subsection{General formulation}
\label{sec:theory}

Consider a single-electron system whose time-evolution is governed by the Hamiltonian,
\begin{equation}
\op{H}(t)=\op{H}_0+\op{V}+\op{H}_I(t),
\label{pd1}
\end{equation}
where $\op{H}_0$ is the free-particle Hamiltonian, $\op{V}$ corresponds to the static interaction, and $\op{H}_I(t)$ accounts for the interaction with 
the laser field, which is always assumed to act for a finite time $T_{\rm p}$, i.e., $\op{H}_I(t)$ vanishes for $t<0$ and $t>T_{\mathrm{p}}$. 
Here, we also define the atomic Hamiltonian,
\begin{equation}
\op{H}_A=\op{H}_0+\op{V},
\label{pd2}
\end{equation}
and the so-called Volkov Hamiltonian,
\begin{equation}
\op{H}_V(t)=\op{H}_0+\op{H}_I(t).
\label{pd3}
\end{equation}
For these three Hamiltonians we introduce the evolution operators,
\begin{align}
\op{U}(t,t')=&\hat{\mathcal{T}}\exp\Bigl( -\ii\int_{t'}^t \dd\tau \op{H}(\tau) \Bigr) , \nonumber \\
\op{U}_A(t,t')=&\ee^{-\ii\op{H}_A (t-t')} , \nonumber \\
\op{U}_V(t,t')=&\hat{\mathcal{T}}\exp\Bigl( -\ii\int_{t'}^t \dd\tau \op{H}_V(\tau) \Bigr) ,
\label{pd4}
\end{align}
where $\hat{\mathcal{T}}$ is the time-ordering operator.

We assume that the atomic Hamiltonian $\op{H}_A$ has both discrete and continuous eigenenergies such that
\begin{equation}
\op{H}_A\ket{B}=E_B\ket{B}, \, 
\op{H}_A\ket{\bm{p};-}=E_{\bm{p}}\ket{\bm{p};-},
\label{pd5}
\end{equation}
where ${\bm p}$ is the asymptotic momentum of the electron. Because the corresponding eigenstates $\ket{B}$ and $\ket{{\bm p};-}$ fulfill the relations
\begin{align}
&\scal{B'}{B}=\delta_{B,B'}, \, \scal{B}{\bm{p};-}=0, \nonumber \\
&\scal{\bm{p}';-}{\bm{p};-}=(2\pi)^3\delta^{(3)}(\bm{p}-\bm{p}'),
\label{pd6}
\end{align}
we can write that
\begin{equation}
\sum_B\ket{B}\bra{B}+\int\frac{\dd^3p}{(2\pi)^3}\ket{\bm{p};-}\bra{\bm{p};-}=\op{I}.
\label{pd7}
\end{equation}
Now, in order to describe ionization, one typically calculates the transition probability amplitude from a bound state $\ket{B}$ to a scattering state $\ket{\bm{p};-}$, 
\begin{equation}
\mathcal{A}_B(\bm{p};-)=\bra{\bm{p};-}\op{S}\ket{B}=\lim_{t\rightarrow\infty}\lim_{t'\rightarrow -\infty}\mathcal{A}_B(\bm{p};t,t'),
\label{pd8}
\end{equation}
where $\op{S}=\op{U}(\infty,-\infty)$ and
\begin{equation}
\mathcal{A}_B(\bm{p};t,t')=\bra{\bm{p};-}\op{U}(t,t')\ket{B}.
\label{pd9}
\end{equation}
Using here the Lippmann-Schwinger equation,
\begin{equation}
\op{U}(t,t')=\op{U}_A(t,t')-\ii\int\dd\tau\op{U}(t,\tau)\op{H}_I(\tau)\op{U}_A(\tau,t'),
\label{pd12}
\end{equation}
and the property $\bra{\bm{p};-}\op{U}_A(t,t')\ket{B}=0$, we arrive at the following expression for the ionization probability amplitude,
\begin{equation}
\mathcal{A}_B(\bm{p};-)=-\ii\int_0^{T_{\mathrm{p}}}\dd t\int\dd^3 x \bigl[\Psi^{(-)}_{\bm{p}}(\bm{x},t)\bigr]^{*}\op{H}_I(t)\psi_B(\bm{x},t),
\label{pd13}
\end{equation}
where $\psi_B(\bm{x},t)=\ee^{-\ii E_B t}\scal{\bm{x}}{B}$ and
\begin{equation}
\bigl[\Psi^{(-)}_{\bm{p}}(\bm{x},t)\bigr]^{*}=\scal{\Psi^{(-)}_{\bm{p}}(t)}{\bm{x}}=\bra{\bm{p};-}\op{U}(T_{\mathrm{p}},t)\ket{\bm{x}}.
\label{pd14}
\end{equation}
Here, we emphasize that the state $\ket{\Psi^{(-)}_{\bm{p}}(t)}$ satisfies the Schr\"odinger equation with the full Hamiltonian $\op{H}(t)$.
Finally, the total probability of ionization equals
\begin{equation}
\mathcal{P}_B=\int\frac{\dd^3p}{(2\pi)^3}|\mathcal{A}_B(\bm{p};-)|^2,
\label{pd10}
\end{equation}
and its momentum distribution is
\begin{equation}
\frac{\dd^3\mathcal{P}_B}{\dd^3p}\equiv\mathcal{P}_B(\bm{p};-)=\frac{1}{(2\pi)^3}|\mathcal{A}_B(\bm{p};-)|^2.
\label{pd11}
\end{equation}

Similarly, the probability amplitude for ionization from the bound state $\ket{B}$ to the final vortex state $\ket{p_{\|},p_{\bot},m;-}$ is defined as
\begin{align}
\mathcal{A}_B(p_{\|},p_{\bot},&m;-)=\bra{p_{\|},p_{\bot},m;-}\op{S}\ket{B} \label{pd8t}\\
=&\frac{1}{2\pi}\int_0^{2\pi}\dd\varphi\, \ee^{-\ii m\varphi}\bra{\bm{p}_{\rm T}(\varphi);-}\op{S}\ket{B} \nonumber \\
=& \frac{1}{2\pi}\int_0^{2\pi}\dd\varphi\, \ee^{-\ii m\varphi}\mathcal{A}_B(\bm{p}_{\rm T}(\varphi);-), \nonumber
\end{align}
or,
\begin{equation}
\mathcal{A}_B(\bm{p}_{\rm T}(\varphi);-)=\sum_{m=-\infty}^{\infty}\ee^{\ii m\varphi}\mathcal{A}_B(p_{\|},p_{\bot},m;-),
\label{pd8ta}
\end{equation}
which follows from the previous section. Hence, the probability distribution of ionization resulting in generation of EVS can be defined as
\begin{equation}
\frac{\dd^2\mathcal{P}_{B,m}}{\dd p_{\|}\dd p_{\bot}}\equiv\mathcal{P}_{B,m}(p_{\|},p_{\bot};-)=\frac{p_{\bot}}{(2\pi)^2}|\mathcal{A}_B(p_{\|},p_{\bot},m;-)|^2.
\label{pd11t}
\end{equation}
Note that this is the most general nonrelativistic description of ionization. 
We will show next that, for the parameters used in this paper, relativistic corrections play already a role and have to be incorporated
into the nonrelativistic theory.

\subsection{Corrected quasi-relativistic SFA}
\label{sec:CSFA}

Since recent experimental~\cite{PressExp} and theoretical~\cite{KKpress,Bandrauk1,Bandrauk2,Bandrauk3,TD2012,R2013,HLH2017,Ivan} investigations, it has become clear that, 
for near infrared pulses of intensities of the order of $10^{14}$~W/cm$^2$ or larger, the effects related to the radiation pressure~\cite{Lebiediew} 
can be detected in photoionization spectra. These effects are accounted for in the relativistic theories based on the Dirac or Klein-Gordon equations. Comparisons 
between the relativistic Dirac and nonrelativistic Schr\"odinger approaches show how the latter has to be modified in order to obtain a good agreement with the relativistic 
treatment for intensities up to $10^{16}$~W/cm$^2$~\cite{KKpress}. This goal can be achieved using the quasi-relativistic strong-field approximation which 
for free-free transitions in intense laser fields has been considered by Ehlotzky~\cite{Ehlo} (see, also~\cite{EJK1998}), whereas for bound-free transitions
by Krajewska and Kami\'nski~\cite{KKpress}. Below, we outline briefly the key ingredients of the corrected (as compared to~\cite{KKpress}) quasi-relativistic strong-field approximation (QRSFA),
which is necessary in the regime of parameters used in this paper.

Generally speaking, the strong-field approximation (SFA) is applicable for high-energy ionization if the kinetic energy of photoelectrons is much larger than the ionization potential of the initial 
bound state, $E_{\mathrm{kin}}(\bm{p})\gg |E_B|$. This condition is very well satisfied in our paper. In such case,
it is justified to expand the full scattering state $\ket{\Psi^{(-)}_{\bm{p}}(t)}$ [Eq.~\eqref{pd14}] in a Born series with respect to the binding potential and, in its lowest order,
to approximate this state by the Volkov solution, $\ket{\psi_{\bm p}^{(0)}(t)}$~\cite{Volkov,VolkovRev1,VolkovRev2}. The latter has a different form, depending on the 
framework we use.

\subsubsection{Relativistic corrections}
\label{qrsfa}

Following Ref.~\cite{KKpress}, we assume that in the QRSFA the interaction Hamiltonian $\op{H}_I(t)$, in the velocity gauge, is
\begin{equation}
\op{H}_I(t)=-\frac{e}{m_{\mathrm{e}}}\bm{A}(\phi)\cdot\opb{p}+\frac{e^2}{2m_{\mathrm{e}}}\bm{A}^2(\phi),
\label{pd19}
\end{equation}
where ${\bm A}(\phi)$ is the vector potential describing the laser pulse with a phase $\phi=\omega t-{\bm k}\cdot{\bm x}$. Here, we introduce 
the fundamental frequency of field oscillations $\omega$ that is related to the pulse duration $T_{\rm p}$ such that $\omega=2\pi/T_{\rm p}$.
The wave vector ${\bm k}$ is defined as ${\bm k}=(\omega/c)\bm{n}$ with a unit vector $\bm{n}$ determining the propagation direction of the laser pulse. 
As stated before, the electromagnetic potential vanishes outside the interval $0 < \phi <2\pi$. Having specified $\op{H}_I(t)$, we know the 
exact form of the Volkov Hamiltonian~\eqref{pd3} and, hence, also the Volkov evolution operator, $\hat{U}_V(t,0)$. 

The Volkov state $\ket{\psi^{(0)}_{\bm{p}}(t)}$ originates from the free-electron state $\ket{\bm{p}}$ which evolves in time in the presence 
of a laser pulse, meaning that
\begin{equation}
\ket{\psi^{(0)}_{\bm{p}}(t)}=\op{U}_V(t,0)\ket{\bm{p}}.
\label{pd18}
\end{equation}
As a result, we obtain the Volkov wave function,
\begin{align}
\psi^{(0)}_{\bm{p}}(\bm{x},t)=\exp\Bigl[&-\ii E_{\mathrm{kin}}(\bm{p})t+\ii\bm{p}\cdot\bm{x} \label{pd20} \\
&+\ii\int_0^{\phi}\dd\phi'\Bigl(\frac{e\bm{A}(\phi')\cdot \bm{p}}{N(\bm{p},\bm{k})}
-\frac{e^2\bm{A}^2(\phi')}{2N(\bm{p},\bm{k})}\Bigr)\Bigr]. \nonumber
\end{align}
Note that for the nonrelativistic theory and the dipole approximation: 
$\phi=\omega t$, $N(\bm{p},\bm{k})=\omega m_{\mathrm{e}}$, $E_{\rm kin}({\bm p})\equiv E_{\rm kin}^{(0)}({\bm p})={\bm p}^2/(2m_{\rm e})$, and the function $\psi^{(0)}_{\bm{p}}(\bm{x},t)$ in~\eqref{pd20} is the exact solution of the 
Schr\"odinger equation. Its generalization, the way it was introduced in~\cite{KKpress}, accounts for two relativistic corrections referred to as the retardation and 
recoil corrections. While we recapture below the essence of these modifications, a new aspect of our approach is to account for the relativistic mass corrections.

The {\it retardation correction}, stating that $\phi=\omega t-{\bm k}\cdot{\bm x}$, reflects the fact that the laser pulse is a propagating wave. 
It follows from~\cite{KKpress} that for near infrared laser fields of intensities up to $10^{16}$~W/cm$^2$, this correction is negligibly small and can be neglected 
in our further analysis. Hence, we shall assume that $\phi\approx\omega t$ in Eq.~\eqref{pd20}. 
The {\it recoil corrections} account for the recoil of the electron during the exchange of momenta with the laser photons, meaning that
\begin{equation}
N(\bm{p},\bm{k})=p\cdot k=\frac{\omega}{c}(\sqrt{\bm{p}^2+(m_{\mathrm{e}}c)^2}-\bm{p}\cdot\bm{n}).
\label{pd22}
\end{equation}
Note that in the nonrelativistic limit: $N(\bm{p},\bm{k})\approx \omega m_{\rm e}$, or if further terms of the nonrelativistic expansion of~\eqref{pd22} are considered~\cite{Nordsieck},
\begin{equation}
N(\bm{p},\bm{k})\approx \omega m_{\mathrm{e}}\Bigl(1-\frac{1}{m_{\mathrm{e}}c}\bm{p}\cdot\bm{n} \Bigr).
\label{pd21}
\end{equation}
This modification of the nonrelativistic Volkov wave function, if compared with the relativistic SFA, is sufficient in describing the radiation pressure effects for 
intensities up to $10^{15}$~W/cm$^2$. It fails, however, for larger intensities~\cite{KKpress}. For this reason, we shall keep in the following
$N({\bm p},{\bm k})$ as defined in~\eqref{pd22} (see, Appendix~\ref{rsfa}). Note that the momentum of the parent ion is also changed during the ionization process. 
However, due to its large mass, it is commonly assumed that this correction only marginally modifies the probability distributions of photoelectrons, 
although it contributes to the overall momentum balance~\cite{Bandrauk1,Bandrauk2,Bandrauk3}.

It appears that the {\it relativistic mass corrections} start to significantly influence ionization for near infrared laser fields  and intensities larger than 
$10^{15}$~W/cm$^2$. It follows from the Klein-Gordon or Dirac equations that the electron kinetic energy, $E_{\mathrm{kin}}(\bm{p})$, is equal to
\begin{align}
E_{\mathrm{kin}}(\bm{p})&=\sqrt{(m_{\mathrm{e}}c^2)^2+(c\bm{p})^2}-m_{\mathrm{e}}c^2 \label{pd24} \\
&\approx \frac{\bm{p}^2}{2m_{\mathrm{e}}}-\frac{\bm{p}^4}{8m_{\mathrm{e}}^3c^2} \, \dots .\nonumber
\end{align}
Keeping this in mind, we ask the question: When can we neglect the higher mass corrections in the Volkov wave~\eqref{pd20}? Since $E_{\mathrm{kin}}(\bm{p})t$ appears 
there in the phase, the nonrelativistic approximation is acceptable if
\begin{equation}
\frac{[E^{(0)}_{\mathrm{kin}}(\bm{p})]^2}{2m_{\mathrm{e}}c^2}T < \pi,
\label{pd25}
\end{equation}
where $E^{(0)}_{\mathrm{kin}}(\bm{p})$ is the nonrelativistic kinetic energy of the photoelectron introduced before and $T$ is a characteristic time 
of the electron-laser-field interaction. For long pulses, we can assume that this time equals the duration of a single cycle, 
$T=2\pi/\omega_{\mathrm{L}}$, where $\omega_{\mathrm{L}}$ is the laser carrier frequency. For short pulses, $T$ denotes the pulse duration $T_{\rm p}$. 
Since these two times are comparable, in our rough estimate we will choose the former one. Hence, the nonrelativistic approximation for the kinetic energy 
of photoelectrons is applicable if
\begin{equation}
\frac{[E^{(0)}_{\mathrm{kin}}(\bm{p})]^2}{m_{\mathrm{e}}c^2\omega_{\mathrm{L}}} < 1.
\label{pd26}
\end{equation}
Specifically, for the Ti:Sapphire laser, the nonrelativistic theory brakes down when the kinetic energy of photoelectrons is at least
\begin{equation}
\sqrt{m_{\mathrm{e}}c^2\omega_{\mathrm{L}}}\approx 860\,\mathrm{eV}.
\label{pd27}
\end{equation}
While this estimate seems to be independent of the laser field intensity, for intensities not exceeding $10^{15}\,\mathrm{W/cm}^2$ the probability of detecting 
such energetic photoelectrons is extremely small. In this case, our estimate has no practical importance. With increasing intensity, however, the high energy portion 
of the spectrum contributes more significantly to the overall ionization probability, as shown, for instance, in Refs.~\cite{KKsuper,CKKsuper,KKcomb,KCKspiral1,KCKspiral2}. 
This situation will be analyzed closely in our numerical simulations, where the full relativistic kinetic energy will be accounted for.

\subsubsection{Probability amplitude of ionization}
\label{sec:amplitudes}

It follows from the above definitions that the probability amplitude of ionization~\eqref{pd13} in the lowest-order 
Born approximation with respect to the final electron state, denoted now as $\mathcal{A}(\bm{p})$, is
\begin{equation}
\mathcal{A}(\bm{p})=-\ii\int_{-\infty}^{\infty}\dd t\int\dd^3x \,\ee^{-\ii\bm{p}\cdot\bm{x}+\ii G(\omega t,\bm{p})}\op{H}_I(t)\psi_B(\bm{x}), 
\label{pad8}
\end{equation}
where
\begin{align}
G(\phi,\bm{p})=\int_0^{\phi}\dd\phi'\Bigl(&\frac{E_{\mathrm{kin}}(\bm{p})-E_B}{\omega} \nonumber \\
&-\frac{e\bm{A}(\phi')\cdot \bm{p}}{N(\bm{p},\bm{k})}+\frac{e^2\bm{A}^2(\phi')}{2N(\bm{p},\bm{k})}\Bigr)
\label{pad8a}
\end{align}
and $\psi_B(\bm{x})=\scal{\bm{x}}{B}$ is the bound state wave function of energy $E_B$, which follows from the Schr\"odinger equation. As stated above, while the 
retardation corrections are neglected in~\eqref{pad8} and~\eqref{pad8a}, the recoil corrections are fully accounted for by taking $N({\bm p},{\bm k})$ defined in Eq.~\eqref{pd22}. 
We will demonstrate later on that, for the considered parameters, $E_{\rm kin}({\bm p})$ has to be treated relativistically, according to~\eqref{pd24}.
This actually follows from the relativistic formulation of the SFA which, for convenience of the reader, is presented in Appendix~\ref{rsfa}.

\subsection{Model}
\label{model}

In our model, the laser pulse is described by the electric field $\bm{\mathcal{E}}(\phi)$,
\begin{equation}
\bm{\mathcal{E}}(\phi)=F_1(\phi)\bm{\varepsilon}_1+F_2(\phi)\bm{\varepsilon}_2,
\label{pad3}
\end{equation}
where two real polarization vectors $\bm{\varepsilon}_1$ and $\bm{\varepsilon}_2$ 
fulfill the relation $\bm{n}=\bm{\varepsilon}_1 \times\bm{\varepsilon}_2 $. As already stated, the pulse lasts for time $T_{\mathrm{p}}$ and, hence, 
$\omega=2\pi/T_{\mathrm{p}}$. The two real functions $F_j(\phi)$ ($j=1,2$) determine the shape of the pulse in the plane-wave front 
approximation~\cite{Neville} such that
\begin{equation}
F_j(\phi)=\mathcal{N}\omega\sin^2\Bigl(\frac{\phi}{2}\Bigr)\sin(N_{\mathrm{osc}}\phi+\delta_j)\cos(\delta+\delta_j)
\label{pad5}
\end{equation}
for $\phi\in [0,2\pi]$ and 0 otherwise. Here, the real constant $\mathcal{N}$ determines the time-averaged intensity of the laser pulse (cf. Ref.~\cite{KKsuper} for details). 
The polarization properties of the field are controlled by the angles $\delta_j$ and $\delta$. We choose in the following: $\delta_j=(j-1)\pi/2$ and $\delta=\pi/4$
for a circularly polarized laser light. The number of cycles is denoted by $N_{\rm osc}$, which allows us to define the laser carrier frequency, $\omega_{\rm L}=N_{\rm osc}\omega$. 
As it follows from Eq.~\eqref{pad3}, the vector potential has the form
\begin{equation}
\bm{A}(\phi)=f_1(\phi)\bm{\varepsilon}_1+f_2(\phi)\bm{\varepsilon}_2,
\label{pad6}
\end{equation}
with
\begin{equation}
f_j(\phi)=-\int_0^{\phi}\dd\phi'\, F_j(\phi'),
\label{pad7}
\end{equation}
and it vanishes for $\phi<0$ and $\phi>2\pi$. 

We use the above model to describe a circularly polarized Ti:Sapphire laser pulse, with the laser carrier frequency $\omega_{\rm L}=1.5498$~eV (wavelength $\lambda=800$~nm). 
While in the following we assume that the pulse consists of three cycles ($N_{\rm osc}=3$), we want to emphasize that we arrive at the same general conclusions for other short 
pulse durations. Such short pulses can be generated experimentally as reported, for instance, in Ref.~\cite{Hung2016}. Moreover, as presented in the captions of 
the figures, we will consider the time-averaged intensities of the order of $10^{16}$~W/cm$^2$.

Our numerical illustrations will concern ionization of a helium ion He$^+$ (i.e., $Z=2$) in the ground state. As it follows from the Dirac equation,
the binding energy of such state is $E_B^{\rm rel}=m_{\mathrm{e}}c^2\sqrt{1-Z^2\alpha^2}$, where $\alpha\approx 1/137$ is the fine-structure constant. 
When taking the nonrelativistic limit, we obtain
\begin{align}
E_B^{\rm rel}-m_{\rm e}c^2&=m_{\mathrm{e}}c^2(\sqrt{1-Z^2\alpha^2}-1) \nonumber \\
&\approx -\frac{1}{2}Z^2\alpha^2m_{\mathrm{e}}c^2-\frac{1}{8}Z^4\alpha^4m_{\mathrm{e}}c^2\,\dots ,
\label{pad1}
\end{align}
where the lowest order term in $\alpha$ corresponds to the nonrelativistic ground state energy of a hydrogen-like ion, the way it follows 
from the Schr\"odinger equation (for He$^+$, $|E_B|=\frac{1}{2}Z^2\alpha^2m_{\mathrm{e}}c^2\approx 54$~eV). Let us note that for He$^+$ and for the Ti:Sapphire laser field,
\begin{equation}
\frac{Z^4\alpha^4m_{\mathrm{e}}c^2}{8\omega_{\mathrm{L}}}\approx 2\times 10^{-3}\ll 1.
\label{pad2}
\end{equation}
Thus, we can neglect the relativistic corrections to the binding energy in our QRSFA (see, Appendix~\ref{rsfa}). However, for heavier ions ($Z\gtrsim10$), this assumption is questionable 
and it becomes necessary to treat the ionization from the ground state in the relativistic framework.

\subsection{Comparison between different approximations}
\label{sec:comparison}

According to the general theory presented in Sec.~\ref{sec:theory}, the total probability of ionization in the QRSFA is 
\begin{equation}
\mathcal{P}_{\mathrm{ion}}=\int\frac{\dd^3p}{(2\pi)^3} |\mathcal{A}(\bm{p})|^2,
\label{pad8aa}
\end{equation}
where ${\cal A}({\bm p})$ is given by~\eqref{pad8}. In the following, we consider two versions of this equation. When in Eq.~\eqref{pad8a},
\begin{itemize}
\item[i)]
there is no mass corrections,
\begin{equation}
E_{\mathrm{kin}}(\bm{p})\approx\frac{\bm{p}^2}{2m_{\mathrm{e}}},
\label{pad10}
\end{equation}
\item[ii)]
mass corrections are fully accounted for,
\begin{equation}
E_{\mathrm{kin}}(\bm{p})=\sqrt{(m_{\mathrm{e}}c^2)^2+(c\bm{p})^2}-m_{\mathrm{e}}c^2.
\label{pad11}
\end{equation}
\end{itemize}
Despite these substitutions in~\eqref{pad8a}, in both cases we define the triply-differential probability distribution as
\begin{equation}
\frac{\dd^3\mathcal{P}}{\dd E_{\mathrm{kin}}\dd^2\Omega_{\bm{p}}}=\frac{m_{\mathrm{e}}|\bm{p}|}{(2\pi)^3} |\mathcal{A}(\bm{p})|^2,
\label{pad8b}
\end{equation}
or, if expressed in atomic units,
\begin{equation}
\mathcal{P}(\bm{p})=\alpha^2m_{\mathrm{e}}c^2 \frac{\dd^3{\cal P}}{\dd E_{\mathrm{kin}}\dd^2\Omega_{\bm{p}}}.
\label{pad8c}
\end{equation}
Note also that the recoil corrections are fully accounted for in both these quasi-relativistic approaches.

\begin{figure}
\includegraphics[width=7cm]{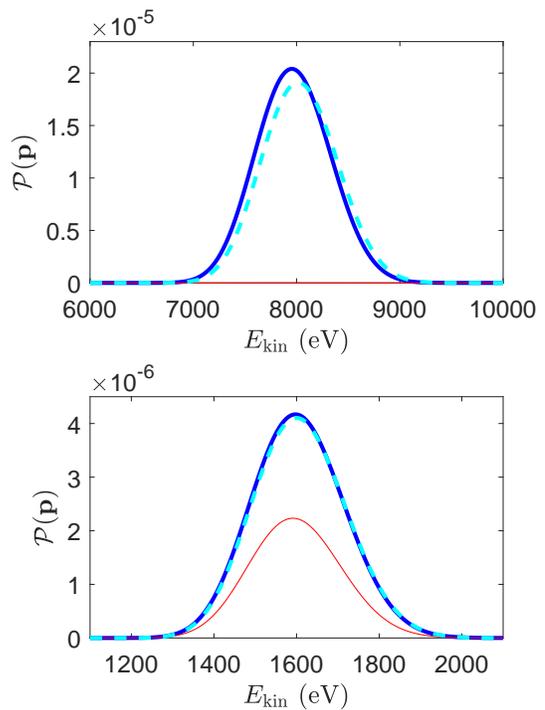}
\caption{Energy probability distributions of ionized electrons, in atomic units, calculated from different theories: the relativistic SFA (thick solid blue line),
the QRSFA accounting only for the recoil corrections (thin solid red line), and the QRSFA accounting for both 
the recoil and mass corrections (thick dashed cyan line). In the upper panel, we present the results for the time-averaged intensity of 
the laser pulse $5\times 10^{16}$~W/cm$^2$ and the polar and azimuthal angles of emission: $\theta_{\bm{p}}=0.4719\pi$ and $\varphi_{\bm{p}}=0$, respectively. 
In the lower panel, we plot the same but for the time-averaged intensity of $10^{16}$~W/cm$^2$ and the polar angle $\theta_{\bm{p}}=0.4874\pi$.
}
\label{fiCompare170914}
\end{figure}

These two versions of the QRSFA will be compared with the relativistic treatment based on the Dirac equation. 
Note that the relativistic SFA accounts exactly for the recoil, retardation, and mass corrections. In this approximation, the total probability of ionization is
\begin{equation}
\mathcal{P}_{\mathrm{ion}}=\frac{1}{2}\sum_{\lambda,\lambda_{\rm i}=\pm}\int\frac{\dd^3p}{(2\pi)^3} |\mathcal{A}_{\lambda_{\rm i}\lambda}(\bm{p})|^2,
\label{prob}
\end{equation}
where we have summed over the final and averaged over the initial electron spin states, $\lambda_{\rm i}$ and $\lambda$, respectively. 
Here, $\mathcal{A}_{\lambda_{\rm i}\lambda}(\bm{p})$ is given in Appendix~\ref{rsfa} [Eq.~\eqref{ampDirac}]. Based on~\eqref{prob}, 
we define the spin-independent triply-differential probability distribution of ionization,
\begin{equation}
\frac{\dd^3\mathcal{P}}{\dd E_{\bm p}\dd^2\Omega_{\bm{p}}}=\frac{m_{\mathrm{e}}|\bm{p}|}{2(2\pi)^3}\sum_{\lambda,\lambda_{\rm i}=\pm} 
|\tilde{\mathcal{A}}_{\lambda_{\rm i}\lambda}(\bm{p})|^2,
\label{probnew}
\end{equation}
with $\displaystyle\tilde{\cal A}_{\lambda_{\rm i}\lambda}({\bm p})={\cal A}_{\lambda_{\rm i}\lambda}({\bm p})\sqrt{\frac{E_{\bm p}}{m_{\rm e}c^2}}$. When expressed in atomic units,
\begin{equation}
\mathcal{P}(\bm{p})=\alpha^2m_{\mathrm{e}}c^2 \frac{\dd^3 {\cal P}}{\dd E_{\bm p}\dd^2\Omega_{\bm{p}}},
\label{probnewnew}
\end{equation}
it represents the quantity to be compared with~\eqref{pad8c}.

In Fig.~\ref{fiCompare170914}, we compare the high-energy spectra of photoelectrons when calculated from either the relativistic SFA (thick solid blue line)
or the QRSFA without the mass corrections (thin solid red line) and fully accounting for them (dashed cyan line). Note also that,
in both quasi-relativistic approaches, the recoil corrections are taken into account.
As expected based on our theoretical analysis, for a three-cycle Ti:Sapphire laser pulse of the nonrelativistic intensity $I=10^{16}\,\mathrm{W/cm}^2$ 
(lower panel), not only the recoil corrections, but also 
the relativistic mass corrections play a significant role in the energy spectra of photoelectrons around 1600~eV. With increasing the 
intensity and the photoelectron kinetic energy (although still nonrelativistic), the role of these corrections become even more important (upper panel). 
By comparing the results derived from the Dirac equation and the quasi-relativistic approach accounting fully for the mass 
corrections, one can conclude that both approaches lead to almost identical distributions. Also, it shows that the effects related 
to the retardation corrections are negligible, which has been already shown in~\cite{KKpress}. Furthermore, for $I=10^{16}\,\mathrm{W/cm}^2$, 
all the considered cases show probability distributions which are qualitatively similar, although their peak values depend on the corrections applied. 
In fact, by scaling all these distributions to their maximum values (i.e., by presenting them in `arbitrary units') one would get nearly identical curves. 
On the other hand, while at intensities close to $5\times10^{16}\,\mathrm{W/cm}^2$ (upper panel) the results accounting for recoil and mass corrections still 
agree very well with the ones obtained from the Dirac theory, this is not the case for the QRSFA neglecting the mass corrections.
It does not only differ considerably but it leads to negligibly small (compared to the full relativistic treatment) probabilities for high-energy ionization.

\section{Generation of vortex states}
\label{sec:twist}

In our further analysis, we will use the QRSFA in which we take into account the recoil and mass corrections fully [i.e., the version ii) above]. 
We have selected this specific approach as, for the laser field intensities and photoelectron kinetic energies considered here,
it very well coincides with the relativistic theory. 

In order to proceed, we write the corrected Volkov wave function \eqref{pd20} in the abstract form
\begin{equation}
\psi^{(0)}_{\bm{p}}(\bm{x},t)=\bra{\bm{x}}\op{U}_{\mathrm{QR-B}}(t)\ket{\bm{p}},
\label{gad1}
\end{equation}
where 
\begin{align}
\label{gad2}
\op{U}_{\mathrm{QR-B}}(t)=&\exp\Bigl[-\ii E_{\mathrm{kin}}(\opb{p})t \\
&+\ii\int_{-\infty}^{\omega t}\dd\phi'\Bigl(\frac{e\bm{A}(\phi')\cdot \opb{p}}{N(\opb{p},\bm{k})}-\frac{e^2\bm{A}^2(\phi')}{2N(\opb{p},\bm{k})}\Bigr)\Bigr] \nonumber
\end{align}
and the integration over $\phi'$ has been extended to $-\infty$ as the vector potential vanishes for $\phi'<0$. 
This allows us to represent the amplitude~\eqref{pad8} in the form~\eqref{tp3}, i.e.,
\begin{equation}
\mathcal{A}(\bm{p})=\bra{\bm{p}}\op{S}_{\mathrm{QR-B}}\ket{B},
\label{gad3}
\end{equation}
where
\begin{equation}
\op{S}_{\mathrm{QR-B}}=-\ii \int_{-\infty}^{\infty}\dd t \,\op{U}_{\mathrm{QR-B}}^{\dagger}(t)\op{H}_I(t)\ee^{-\ii\op{H}_At},
\label{gad4}
\end{equation}
as $\op{H}_I(t)$ vanishes for $t<0$ and $t>T_{\mathrm{p}}$. Thus, we can formally interpret $\op{S}_{\mathrm{QR-B}}$ as the evolution operator for the transition 
from a bound state to the high-energy continuum in the quasi-relativistic and Born approximations. This also shows that the probability amplitudes of ionization 
into vortex states can be calculated from $\mathcal{A}(\bm{p})$ by the Fourier decomposition \eqref{tw11a},
\begin{align}
\mathcal{A}(\bm{p}_{\mathrm{T}}(\varphi))&=\sum_{m=-\infty}^{\infty}\ee^{\ii m\varphi}\bra{p_{\|},p_\perp,m}\op{S}_{\mathrm{QR-B}}\ket{B}\nonumber\\
&=\sum_{m=-\infty}^{\infty}\ee^{\ii m\varphi}\mathcal{A}_m(p_{\|},p_{\bot}).
\label{gad5}
\end{align}
Here, we have changed the notation from ${\cal A}(p_{\|},p_\perp,m)$ to $\mathcal{A}_m(p_{\|},p_{\bot})$ in order to separate the discrete variable $m$ 
from the remaining two continuous ones, $p_{\|}$ and $p_{\bot}$.

\subsection{Ionization spiral}
\label{sec:ionspiral}

\begin{figure}
\includegraphics[width=5cm]{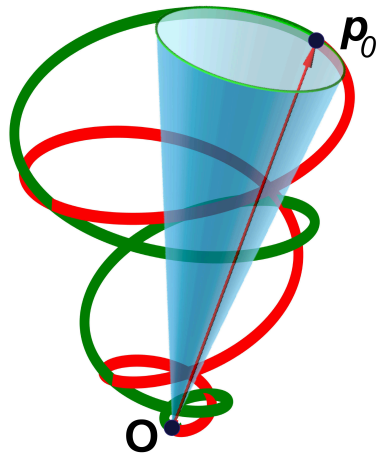}
\caption{Schematic representation of the kinematics in momentum space considered in this paper. The thick line represents the ionization spiral 
$\bm{p}_{\mathrm{S}}(\phi)$ with the red (lighter) line corresponding to the ramp-up portion of the laser pulse, and the dark green (darker) line
to the ramp-down portion. The twisted momentum, $\bm{p}_{\mathrm{T}}(\varphi)$, rotates on the surface of the semitransparent blue cone such that, for a particular value of $\varphi=\varphi_0$, it touches the ionization spiral, i.e., there exists a phase $\phi=\phi_0$ such that $\bm{p}_{\mathrm{S}}(\phi_0)=\bm{p}_{\mathrm{T}}(\varphi_0)=\bm{p}_0$. In our analysis, we choose $\phi_0=\pi$ and $\varphi_0=0$. 
Note that, for visual purposes, the vertical and horizontal axes are not in scale.
}
\label{spiraltouchp}
\end{figure}

As it has been shown in~\cite{KCKspiral1,KCKspiral2}, the high-energy ionization is unlikely unless the photoelectron momentum $\bm{p}$ approaches ${\bm p}_{\rm S}(\phi)$,
which is parametrized by the laser phase $\phi\in[0,2\pi]$ such that
\begin{equation}
\bm{p}^{\bot}_{\rm S}(\phi)=e\bm{A}(\phi), \quad p^{\|}_{\rm S}(\phi)=\frac{e^2\bm{A}^2(\phi)}{2m_{\mathrm{e}}c\sqrt{1-Z^2\alpha^2}}.
\label{gad7}
\end{equation}
Here, $\bm{p}^{\bot}_{\rm S}$ and $p^{\|}_{\rm S}$ are the perpendicular and parallel components of momentum ${\bm p}_{\rm S}(\phi)$ with respect to the direction 
of propagation of the laser pulse ${\bm n}$, and have to be distinguished from the cylindrical coordinates introduced in Sec.~\ref{sec:twistfree}. They define
a curve in momentum space,
\begin{equation}
\bm{p}_{\mathrm{S}}(\phi)=\bm{p}^{\bot}_{\rm S}(\phi)+p^{\|}_{\rm S}(\phi)\bm{n},
\label{gad8}
\end{equation}
which we will call the {\it ionization spiral}. Several properties of the ionization probability distribution can be deduced from this analytical prediction~\eqref{gad8}. 
For instance, for the laser pulse parameters considered in Fig.~\ref{fiCompare170914}, $\bm{p}_{\mathrm{S}}(\pi)$ (i.e., the value of ${\bm p}_{\rm S}$ at the pulse
maximum) defines the polar and azimuthal angles ($\theta_{\bm p}$ and $\varphi_{\bm p}$, respectively) at which the ionized electron is detected with the locally largest
probability distribution. These values are presented in the caption of Fig.~\ref{fiCompare170914}. Also, the kinetic energy corresponding to $\bm{p}_{\mathrm{S}}(\pi)$ determines the central energy 
of the probability distribution, i.e., the energy at which the distribution is peaked. Note, however, that the predictions arising from the momentum spiral are valid 
only for the high-energy ionization (i.e., for sufficiently intense pulses)~\cite{KCKspiral1,KCKspiral2}. Thus, even though we define 
$\bm{p}_{\mathrm{S}}(\phi)$ for all possible laser phases $\phi$, its interpretation as photoelectron momentum detected with maximum probability is only valid for 
the high-energy portion of ionization spectrum. Based on our numerical analysis, we can roughly quantify what `the high-energy portion of ionization spectrum' means. 
Namely, it relates to photoelectron kinetic energies larger than $10|E_B|$~\cite{CKKsuper}.

\begin{figure}
\includegraphics[width=6cm]{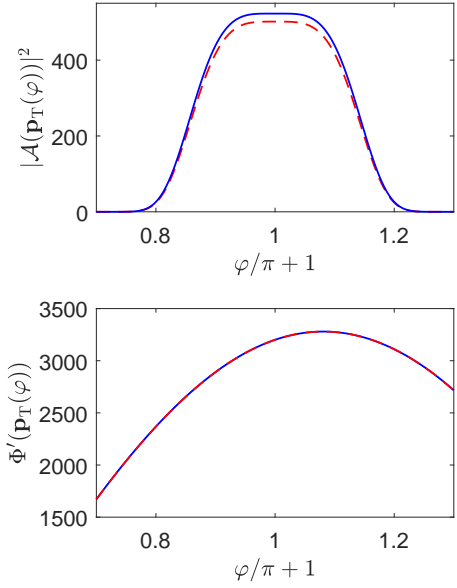}
\caption{Modulus squared of the probability amplitude of ionization ${\cal A}({\bm p}_{\rm T}(\varphi))$, in relativistic units, (upper panel) 
and the derivative of its phase (bottom panel) as functions of the twist angle $\varphi$. While the time-averaged intensity of the laser pulse is 
$5\times 10^{16}\,\mathrm{W/cm}^2$, the remaining parameters of the pulse are specified in Sec.~\ref{model}. In cylindrical coordinates
defined by the angles $\theta_{\mathrm{T}}=0.37\pi$ and $\varphi_{\mathrm{T}}=0$, the photoelectron final momentum is such that $p_{\|}=0.17m_{\mathrm{e}}c$
and $p_{\bot}=0.055m_{\mathrm{e}}c$. In addition, we take: $\phi=\pi$, $\delta p_{\|}=\delta p_{\bot}=\delta\varphi_{\mathrm{T}}=0$, and $\delta\theta_{\mathrm{T}}=-0.1\pi$,
meaning that $\beta_{\rm T}=0.1\pi$. 
The solid blue line represents the results based on the Dirac equation (i.e., the relativistic SFA) with the initial and final electron spin projections on the direction 
of laser pulse propagation. The dashed red line is for the quasi-relativistic approach ii) specified in Sec.~\ref{sec:comparison}. The results presented here are limited to twist angles 
for which the modulus of the probability amplitude is sufficiently different than 0, otherwise, the determination of the phase is erratic.
}
\label{xf5x15amp2phase170920}
\end{figure}
\begin{figure}
\includegraphics[width=6cm]{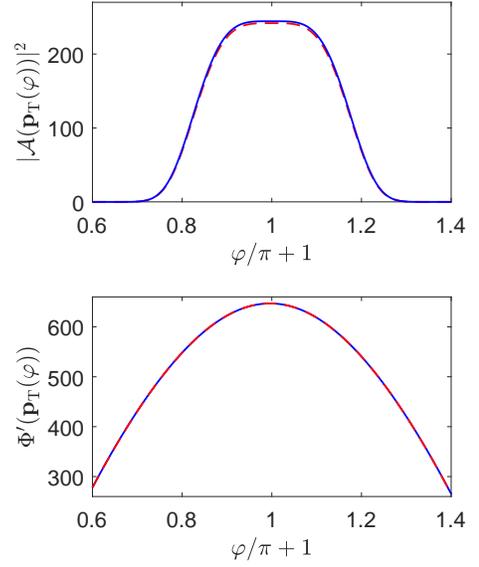}
\caption{The same as in Fig.~\ref{xf5x15amp2phase170920}, but for the laser field intensity $10^{16}\,\mathrm{W/cm}^2$ and for the parameters: 
$p_{\|}=0.075m_{\mathrm{e}}c$, $p_{\bot}=0.024m_{\mathrm{e}}c$, $\theta_{\mathrm{T}}=0.387\pi$, and $\varphi_{\mathrm{T}}=0$. The remaining parameters are still:
$\phi=\pi$, $\delta p_{\|}=\delta p_{\bot}=\delta\varphi_{\mathrm{T}}=0$, $\delta\theta_{\mathrm{T}}=-0.1\pi$, and $\beta_{\rm T}=0.1\pi$. 
}
\label{xf1x16amp2phase170920}
\end{figure}

As we have stated above, in high-energy ionization, the photoelectrons with momenta far away from the spiral~\eqref{gad8} are emitted with very small probabilities. 
Therefore, for an arbitrary choice of twisted momenta $\bm{p}_{\mathrm{T}}(\varphi)$, a very weak ionization signal is expected. A stronger signal will be obtained
only for those momenta $\bm{p}_{\mathrm{T}}(\varphi)$ which, for some values of the twist angle $\varphi$, approach $\bm{p}_{\mathrm{S}}(\phi)$ in momentum space. Now, 
we shall construct such $\bm{p}_{\mathrm{T}}(\varphi)$.

Let us select a particular laser phase $\phi_0$ and define the momentum $\bm{p}_0=\bm{p}_{\mathrm{S}}(\phi_0)$, which points in the direction determined by the polar and 
azimuthal angles $\theta_0$ and $\varphi_0$, respectively. Next, we fix the angles $\theta_{\mathrm{T}}$ and $\phi_{\mathrm{T}}$ as follows
\begin{equation}
\theta_{\mathrm{T}}=\theta_0+\delta\theta_{\mathrm{T}}, \quad \varphi_{\mathrm{T}}=\varphi_0+\delta\varphi_{\mathrm{T}},
\label{gad9}
\end{equation}
with arbitrary increments $\delta\theta_{\mathrm{T}}$ and $\delta\varphi_{\mathrm{T}}$. These two angles ($\theta_{\rm T}$ and $\varphi_{\rm T}$) determine the cylindrical 
coordinates with symmetry axis $\bm{n}_{\|}$ and two perpendicular vectors, $\bm{n}_{\bot,1}$ and $\bm{n}_{\bot,2}$~\eqref{twisttriad}. In this system of coordinates, we have
\begin{equation}
p_{0\|}=\bm{p}_0\cdot\bm{n}_{\|}, \quad p_{0\bot}=\sqrt{\bm{p}_0^2-p_{0\|}^2},
\label{gad10}
\end{equation}
and so, the family of twisted momenta $\bm{p}_{\mathrm{T}}(\varphi)$ is defined,
\begin{align}
\label{gad11}
\bm{p}_{\mathrm{T}}(\varphi)&=(p_{0\|}+\delta p_{\|})\bm{n}_{\|} \\
&+(p_{0\bot}+\delta p_{\bot})(\bm{n}_{\bot,1}\cos\varphi+\zeta_H\bm{n}_{\bot,2}\sin\varphi).
\nonumber
\end{align}
As before, we choose the helicity of the vortex state such that $\zeta_H=1$. In principle, the increments $\delta p_{\|}$ and $\delta p_{\bot}$ can be chosen
arbitrarily. However, they should be close to 0 for the twisted momenta $\bm{p}_{\mathrm{T}}(\varphi)$ to approach the spiral $\bm{p}_{\mathrm{S}}(\phi)$. 
Such a choice of $\bm{p}_{\mathrm{T}}(\varphi)$ is schematically illustrated in Fig.~\ref{spiraltouchp} for $\phi_0=\pi$ (i.e., when both the strength of the laser 
pulse and the length of $\bm{p}_0$ are maximum), $\delta p_{\|}=\delta p_{\bot}=\delta\varphi_{\mathrm{T}}=0$, and for $\delta\theta_{\mathrm{T}}=-0.1\pi$. 
This means that the twisted momenta rotate on a cone with the half-opening angle $\beta_{\rm T}=0.1\pi$.
For these parameters, the curves $\bm{p}_{\mathrm{S}}(\phi)$ and $\bm{p}_{\mathrm{T}}(\varphi)$ are tangent to each other for $\phi=\pi$ and $\varphi=0$.

In Figs.~\ref{xf5x15amp2phase170920} and~\ref{xf1x16amp2phase170920}, we present the modulus squared and the phase derivative of the probability amplitude 
of ionization $\mathcal{A}(\bm{p}_{\mathrm{T}}(\varphi))$ as functions of the twist angle $\varphi$. While Fig.~\ref{xf5x15amp2phase170920} 
relates to a time-averaged laser pulse intensity $I=5\times 10^{16}\,\mathrm{W/cm}^2$, Fig.~\ref{xf1x16amp2phase170920} is obtained for $I=10^{16}\,\mathrm{W/cm}^2$.  
It can be seen in both figures that the probability amplitudes of ionization are large for $\varphi$ close to 0. This is the case discussed above in relation to 
Fig.~\ref{spiraltouchp}, when the twisted momentum $\bm{p}_{\mathrm{T}}(\varphi)$ approaches the ionization spiral. This confirms numerically our earlier hypothesis.
In addition, we observe a significant dependence of the amplitude phase,
\begin{equation}
\Phi(\bm{p}_{\mathrm{T}}(\varphi))=\arg \mathcal{A}(\bm{p}_{\mathrm{T}}(\varphi)),
\label{gad13}
\end{equation}
and its derivative, 
\begin{equation}
\Phi'(\bm{p}_{\mathrm{T}}(\varphi))=\frac{\dd}{\dd \varphi}\Phi(\bm{p}_{\mathrm{T}}(\varphi)),
\label{gad13a}
\end{equation}
on the twist angle $\varphi$. It is also worth noting that Figs.~\ref{xf5x15amp2phase170920} and~\ref{xf1x16amp2phase170920} present the results based on the QRSFA
accounting fully for the mass corrections and electron recoil (dashed red curve) and based on the relativistic SFA (solid blue curve). A very good agreement between both theories
is observed not only for the modulus of the ionization probability amplitudes but also for the amplitude phases~\eqref{gad13}, up to an irrelevant constant term.
Once again we see that our quasi-relativistic approach correctly describes the high-energy ionization in the considered regime of parameters.

\subsection{OAM distributions}
\label{sec:OAM}

In this section, using the quasi-relativistic description, we will analyze probability distributions of generating the EVS carrying large
orbital angular momenta $m$. We will refer to them as the orbital angular momenta (OAM) distributions, $|{\cal A}_m(p_{\|},p_\perp)|^2$, where ${\cal A}_m(p_{\|},p_\perp)$
is implicitly defined in~\eqref{gad5}. 

\begin{figure}
\includegraphics[width=7cm]{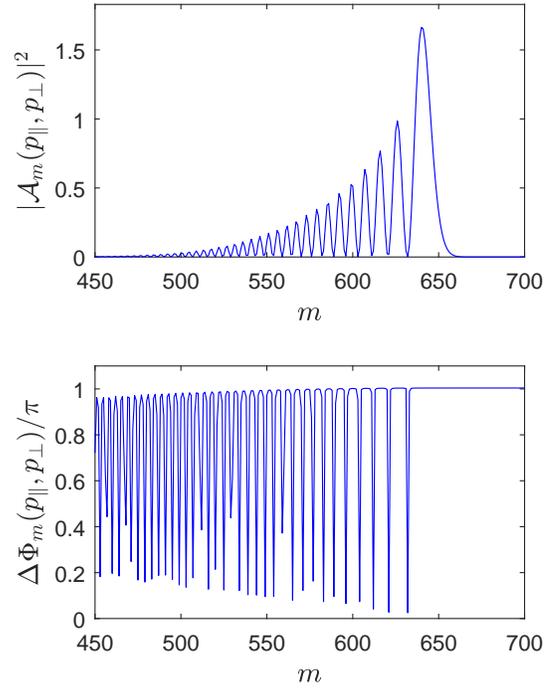}
\caption{The OAM distribution (upper panel), in relativistic units, for the family of vortex states represented in Fig.~\ref{xf1x16amp2phase170920}. The discrete derivative of the phase of the probability amplitude [cf., Eq.~\eqref{gad15}] 
is also presented (lower panel). For visual purposes, in both panels the points corresponding to the integer values of $m$ have been connected by the solid line.
}
\label{xfi1x16oam170920}
\end{figure}
\begin{figure}
\includegraphics[width=7cm]{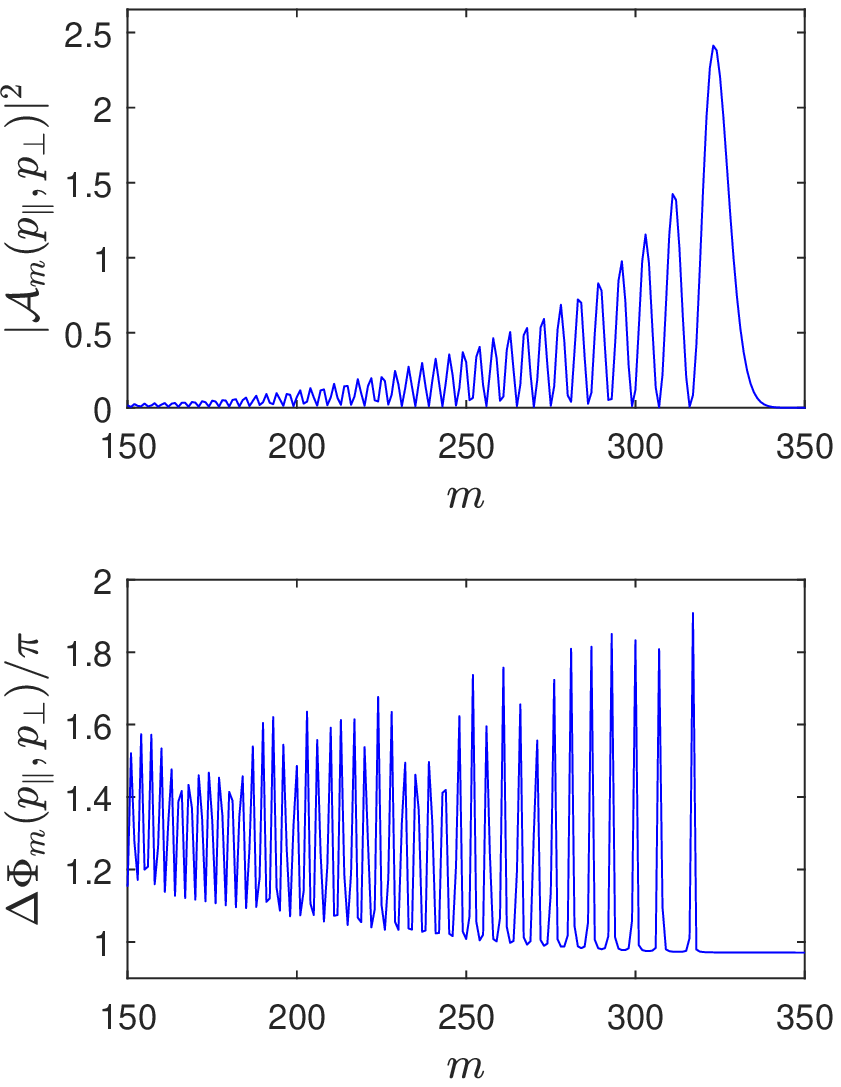}
\caption{The same as in Fig.~\ref{xfi1x16oam170920} but for $p_{\|}=0.078m_{\mathrm{e}}c$, $p_{\bot}=0.012m_{\mathrm{e}}c$, $\theta_{\mathrm{T}}=0.437\pi$, and $\varphi_{\mathrm{T}}=0$.
Moreover, $\phi=\pi$, $\delta p_{\|}=\delta p_{\bot}=\delta\varphi_{\mathrm{T}}=0$, and $\delta\theta_{\mathrm{T}}=-0.05\pi$; i.e., now the opening angle of the cone of twisted 
momenta is two times smaller than in Fig.~\ref{xfi1x16oam170920}. 
}
\label{xfi1x16v1oam170920}
\end{figure}

Some properties of OAM distributions can be anticipated already from Figs.~\ref{xf5x15amp2phase170920} and~\ref{xf1x16amp2phase170920}.  
It follows from these figures that the phase derivative, $\Phi'(\bm{p}_{\mathrm{T}}(\varphi))$, is large. Because of the definition~\eqref{gad5} and general
properties of the Fourier transform, one can conclude that if $\Phi'(\bm{p}_{\mathrm{T}}(\varphi))$ takes large values then the EVS with substantial topological charges $m$ will be generated.
Additionally, since the second and higher derivatives of $\Phi(\bm{p}_{\mathrm{T}}(\varphi))$ are also significantly different from zero (contrary to what is observed 
for the supercontinuum in ionization~\cite{KKsuper}, but similarly to what is predicted for the Compton process~\cite{KK2014a,KK2014b}), we can expect that the OAM distributions
will attain a chirp. This is illustrated in Fig.~\ref{xfi1x16oam170920}.

In the upper panel of Fig.~\ref{xfi1x16oam170920}, we show the discrete OAM distribution, $|\mathcal{A}_m(p_{\|},p_{\bot})|^2$, for the time-averaged 
laser field intensity $10^{16}$~W/cm$^2$. The electron final momenta, calculated in the coordinate system determined by the angles
$\theta_{\mathrm{T}}=0.387\pi$ and $\varphi_{\mathrm{T}}=0$, are: $p_{\|}=0.075\, m_{\mathrm{e}}c$ and $p_{\bot}=0.024\, m_{\mathrm{e}}c$.
Note that in order to obtain the OAM probability distribution out of this figure, one has to multiply $|\mathcal{A}_m(p_{\|},p_{\bot})|^2$
by $p_{\bot}/(2\pi)^2$. Here, we observe a chirp-type structure with the dominant peak centered at $m=645$, which roughly corresponds to the maximum 
value of $\Phi'(\bm{p}_{\mathrm{T}}(\varphi))$ presented in Fig.~\ref{xf1x16amp2phase170920}. Such a coincidence is in full agreement with 
the general property of the Fourier transform, which states that the linear part of the phase is responsible for the `shift' of the Fourier components. 
In our case, this shift occurs towards positive values of $m$. Had we consider the opposite circular polarization of the laser pulse [i.e., $\delta=-\pi/4$
in Eq.~\eqref{pad5}] we would observe an identical shift, but towards negative values. Moreover, for the higher laser pulse intensity $5\times 10^{16}\,\mathrm{W/cm}^2$, 
as expected from the lower panel of Fig.~\ref{xf5x15amp2phase170920}, the probability distribution acquires its maximum values for larger $m$, $m\approx 3200$.

In the lower panel of Fig.~\ref{xfi1x16oam170920}, we plot the discrete derivative of the phase of the probability amplitude,
\begin{equation}
\Phi_m(p_{\|},p_{\bot})=\arg \mathcal{A}_m(p_{\|},p_{\bot}),
\label{gad14}
\end{equation}
defined as
\begin{equation}
\Delta\Phi_m(p_{\|},p_{\bot})=\Phi_m(p_{\|},p_{\bot})-\Phi_{m-1}(p_{\|},p_{\bot}) \mod 2\pi .
\label{gad15}
\end{equation}
Except for particular values of $m$, for which the ionization probability is very small, the phases of $\mathcal{A}_m(p_{\|},p_{\bot})$ increase approximately linearly with $m$, i.e.,
\begin{equation}
\Phi_m(p_{\|},p_{\bot})\approx\Phi_0(p_{\|},p_{\bot})+m\pi \mod 2\pi .
\label{gad16}
\end{equation}
Due to this regularity, the inverse discrete Fourier transform leads to the smooth dependence of $|\mathcal{A}(\bm{p}_{\mathrm{T}}(\varphi))|$ and $\Phi(\bm{p}_{\mathrm{T}}(\varphi))$ on the twist angle $\varphi$.

In Fig.~\ref{xfi1x16v1oam170920}, we show the same as in Fig.~\ref{xfi1x16oam170920} but the cone of twisted momenta is twice that narrow, i.e., a half-opening angle of the cone 
is now $\beta_{\mathrm{T}}=0.05\pi$. Both figures exhibit a similar behavior, except that the probability distribution is now peaked at around two times smaller values 
of $m$, namely, $m\approx 325$. Similar studies carried out for a larger opening angle, with $\beta_{\mathrm{T}}=0.2\pi$, show that the maximum of the OAM distribution
is shifted towards larger values of the topological charge. Specifically, for $\beta_{\rm T}=0.2\pi$ such maximum is found at $m\approx 1120$. This demonstrates that, by changing the angles of 
electron propagation $\theta_{\mathrm{T}}$ and $\varphi_{\mathrm{T}}$, one can select a group of vortex states of topological charges $m$ gathered around a specific value.

\begin{figure}
\includegraphics[width=7.5cm]{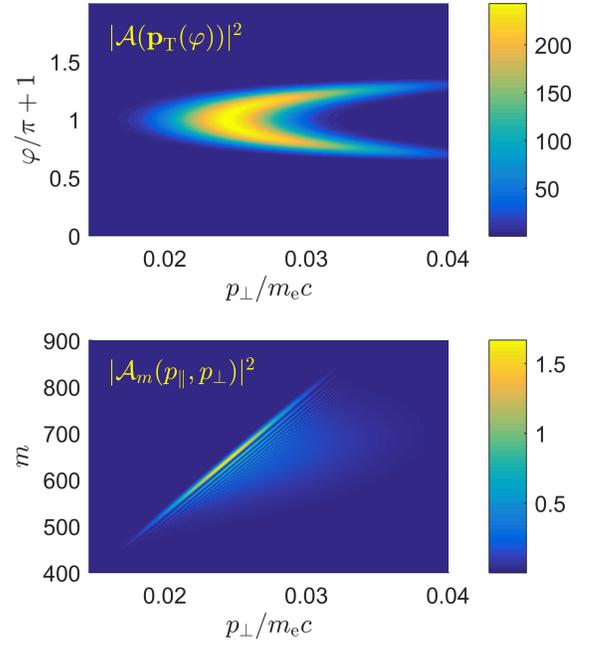}
\caption{Color mappings of ionization probability distributions $|\mathcal{A}(\bm{p}_{\mathrm{T}}(\varphi))|^2$ (upper panel) and $|\mathcal{A}_m(p_{\|},p_{\bot})|^2$ 
(lower panel) for the fixed $p_{\|}=0.075\, m_{\mathrm{e}}c$, and for the polar and azimuthal angles $\theta_{\mathrm{T}}=0.387\pi$ and $\varphi_{\mathrm{T}}=0$. 
The maxima of the OAM distribution (lower panel) depend linearly on the photoelectron perpendicular momentum $p_{\bot}$. Both distributions are for 
$\delta\theta_{\mathrm{T}}=-0.1\pi$ and $\delta\varphi_{\mathrm{T}}=0$.
}
\label{xfiPar2D171002}
\end{figure}

\begin{figure}
\includegraphics[width=7.5cm]{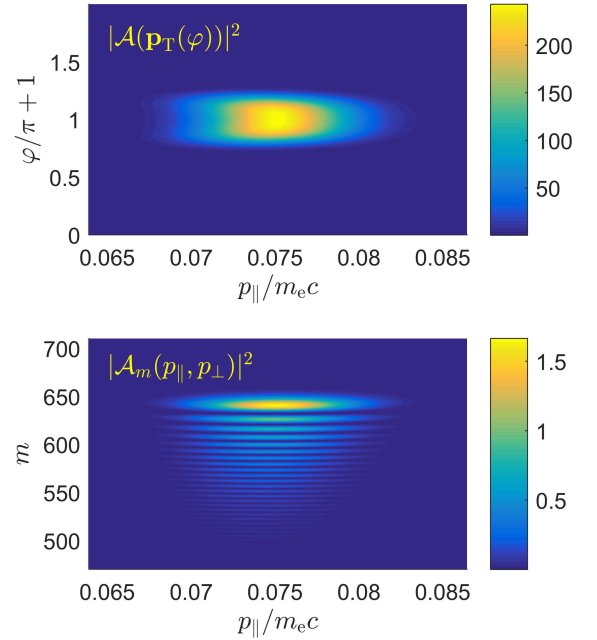}
\caption{The same as in Fig.~\ref{xfiPar2D171002} but as a function of $p_{\|}$ and for the fixed momentum $p_{\bot}=0.024\, m_{\mathrm{e}}c$. 
Note that the maxima of the OAM distribution (lower panel) are independent of the photoelectron parallel momentum $p_{\|}$.
}
\label{xfiPer2D171002}
\end{figure}

The family of twisted momenta~\eqref{gad11} with $\delta p_{\|}=\delta p_{\bot}=0$ represents the most optimal choice for the generation of EVS photoelectron states. 
This is well seen in Figs.~\ref{xfiPar2D171002} and~\ref{xfiPer2D171002}. Note that in both figures we refer to the cylindrical coordinate system
such that $\theta_{\mathrm{T}}=0.387\pi$ and $\varphi_{\mathrm{T}}=0$. Specifically, in the upper panel of Fig.~\ref{xfiPar2D171002} we show the color mapping of the probability
distribution $|{\cal A}({\bm p}_{\rm T}(\varphi))|^2$ as a function of the perpendicular momentum of
the final electron $p_\bot$ and the twist angle $\varphi$. Here, the results are for the fixed value of the electron parallel momentum
$p_{\|}=0.075\, m_{\mathrm{e}}c$. As expected, $|\mathcal{A}(\bm{p}_{\mathrm{T}}(\varphi))|^2$ reaches its maximum value at the twist angle $\varphi=0$ (i.e., when the 
twisted momenta $\bm{p}_{\mathrm{T}}(\varphi)$ touch the ionization spiral $\bm{p}_{\mathrm{S}}(\phi)$ at the pulse maximum, $\phi=\pi$). This happens for 
$p_{\bot}=0.024\, m_{\mathrm{e}}c$, in agreement with the results presented in Fig.~\ref{xf1x16amp2phase170920}. Also, the probability distribution
presented in the upper panel of Fig.~\ref{xfiPer2D171002} peaks at the exact same values.

In the lower panels of Figs.~\ref{xfiPar2D171002} and~\ref{xfiPer2D171002} we show the OAM distributions $|\mathcal{A}_m(p_{\|},p_{\bot})|^2$, which
consist of many parallel stripes. While the ones with the largest topological charge dominate, the sidebands characterized by smaller $m$ gradually disappear. 
If we consider the case of fixed $p_{\|}$ (Fig.~\ref{xfiPar2D171002}), one can observe that the positions of maxima of the distribution change linearly with $p_{\bot}$. 
This is understandable since the orbital angular momentum in the ${\bm n}_{\|}$-direction is a linear function of $p_\bot$, with a slope $x_{\bot}$.
The quantity $x_\bot$ can be interpreted as a perpendicular size of the EVS wave packet. Specifically, based on data plotted in Fig.~\ref{xfiPar2D171002},
we estimate for those vortex states that $x_\bot\approx 10$~nm. On the other hand, by considering the case of fixed $p_{\bot}$ (Fig.~\ref{xfiPer2D171002}), 
the maxima of the OAM distribution are located at specific values of the topological charge, independently of $p_{\|}$. This means that, for the given $m$ and 
$p_{\bot}$, the ionization probability distribution as the function of $p_{\|}$ forms a broad supercontinuum, similar to the one observed for photoelectrons 
with linear momenta~\cite{KKsuper,KKcomb,KCKspiral1,KCKspiral2}. This has a potential to employ such photoelectron wave packets in 5-d electron diffraction. 
Such technique, an extension of the 4-d diffraction which is based on the use of femtosecond electron wave packets (see, e.g., \cite{ESI2000,SM2011,Baum2013,IABPR2014,Oxley2017}), 
would be able to probe helical (or magnetic) properties of matter at different times.

\section{Conclusions}
\label{sec:Conclusions}

We have studied generation of the EVS in ionization by short and intense laser pulses. For this purpose, we have developed a quasi-relativistic approach
going beyond our recent formulation presented in~\cite{KKpress}. As we have shown for near infrared laser pulses and intensities of the order of $10^{16}$~W/cm$^2$,
our modified QRSFA, that accounts for the recoil and mass relativistic corrections, gives quantitatively good results as compared to the relativistic SFA.
We have used this approach to demonstrate that the vortex states of large topological charge (approaching 1000) are generated under current conditions.
It follows from our investigations that such states are detected provided that the family of twisted momenta
approach the ionization spiral. The latter defines the region in momentum space where the ionization occurs with significant probabilities~\cite{KCKspiral1,KCKspiral2}.

We have shown that, for the fixed perpendicular electron momentum $p_{\bot}$ and topological charge $m$, the ionization spectrum form the 
supercontinuum~\cite{KKsuper,KCKspiral1,KCKspiral2}. This means that the EVS might be interesting and important subjects for further studies, 
as they can probe a new degree of freedom (namely, chirality) in electron diffraction experiments. In order to generate few femtosecond or 
attosecond electron vortex wave packets, the creation of photoelectrons of relativistic energies is necessary~\cite{KCKspie}. In this case, however, 
the free-electron states of well defined orbital angular momentum  cannot be defined (see, e.g., Refs.~\cite{BB2017,Barnett2017}). This problem is going 
to be explored in our further investigations.

\section*{Acknowledgements}

This work is supported by the National Science Centre (Poland) under Grant No. 2014/15/B/ST2/02203.

\appendix

\section{Relativistic SFA}
\label{rsfa}

Below, we introduce the relativistic SFA which is based on the Dirac equation (for details, see~\cite{KKsuper}). 
For this purpose, we use the four-vector notation. For two arbitrary four-vectors, $a$ and $b$, we define their scalar product as
$a\cdot b=a_\mu b^\mu$, where the Einstein summation convention is used. We use also the Feynman slash notation for a contraction
with the Dirac gamma matrices, $\slashed{a}=\gamma_\mu a^\mu$, and $\bar{u}=u^\dagger\gamma^0$ for bispinors.

In the relativistic SFA, the interaction Hamiltonian in the velocity gauge is
\begin{equation}
\hat{H}_I(x)=ec\gamma^0\slashed{A}(k\cdot x),
\label{rsfa1}
\end{equation}
where $A(\phi)=(0,{\bm A}(\phi))$ and $\phi=\omega t-{\bm k}\cdot{\bm x}$. Hence, the Volkov solution for an electron embedded in the laser field becomes~\cite{VolkovRev1,VolkovRev2}
\begin{equation}
 \psi_{{\bm p}\lambda}^{(0)}(x)=\sqrt{\frac{m_{\rm e}c^2}{E_{\bm p}}}\Bigl(1-\frac{e}{2k\cdot p}\slashed{A}\slashed{k}\Bigr)u_{{\bm p}\lambda}\ee^{-\ii S_p(x)},
\label{rsfa2}
\end{equation}
with
\begin{equation}
S_p(x)=p\cdot x+\int_0^{k\cdot x}\dd\phi\Bigl[\frac{eA(\phi)\cdot p}{p\cdot k} -\frac{e^2A^2(\phi)}{2p\cdot k}\Bigr]
\label{rsfa3}
\end{equation}
and
\begin{equation}
p=(E_{\bm{p}}/c,\bm{p}), \quad E_{\bm{p}}=\sqrt{(c\bm{p})^2+(m_{\mathrm{e}}c^2)^2}.
\label{rsfa4}
\end{equation}
In Eq.~\eqref{rsfa2}, $u_{\bm{p}\lambda}$ is the free-electron bispinor such that
\begin{equation}
(\slashed{p}-m_{\mathrm{e}}c)u_{\bm{p}\lambda}=0,
\label{rsfa5}
\end{equation}
which satisfies the normalization condition $\bar{u}_{\bm{p}\lambda}u_{\bm{p}\lambda'}=\delta_{\lambda\lambda'}$, with $\lambda=\pm 1$ labeling the spin degrees of freedom.

Without going into details of our calculations, which are presented in~\cite{KKsuper}, we rewrite the probability amplitude of ionization
from the bound state of a hydrogen-like ion into the continuum such that
\begin{align}
{\cal A}_{\lambda_{\rm i}\lambda}({\bm p})=&-\ii\sqrt{\frac{m_{\rm e}c^2}{E_{\bm p}}}\int_{-\infty}^{\infty}\dd t\int\dd^3x \nonumber\\
&\times\ee^{-\ii\bigl({\bm p}+\frac{E_B^{\rm rel}-E_{\bm p}}{c}{\bm n}\bigr)\cdot{\bm x}+\ii G(\omega t-{\bm k}\cdot {\bm x},{\bm p})}\nonumber\\
&\times \bar{u}_{{\bm p}\lambda} \Bigl(1+\frac{e}{2k\cdot p}\slashed{A}\slashed{k}\Bigr)\gamma^0
\hat{H}_I(x)\psi_B({\bm x})
\label{ampDirac}
\end{align}
and
\begin{align}
G(\phi,{\bm p})=\int_0^\phi\dd\phi'\, \Bigl(&\frac{E_{\bm p}-E_B^{\rm rel}}{\omega}\nonumber\\
&-\frac{e{\bm A}(\phi')\cdot {\bm p}}{k\cdot p}+\frac{e^2{\bm A}^2(\phi')}{2k\cdot p}\Bigr).
\label{Gnew}
\end{align}
Note that the relativistic theory takes into account the electron spin. Specifically, in~\eqref{ampDirac}, we have the initial $\lambda_{\rm i}$ and final $\lambda$ 
spin degrees of freedom of the ionized electron. It happens however that, for the parameters considered in this paper, the spin effects
are negligible. In other words, the ionization occurs with no spin flip (taking into account that the spin quantization axis is parallel to the laser pulse propagation
direction). We also need to stress that even though we keep the same notation for the bound state
as in Sec.~\ref{qrsfa}, this time $\psi_B({\bm x})$ is a four-component wave function which follows from the Dirac equation~\cite{CKKsuper}
and depends on $\lambda_{\rm i}$.

To get some insight into the origin of relativistic corrections introduced in Sec.~\ref{qrsfa}, we focus here on the exponent of Eq.~\eqref{ampDirac}. Note that its phase differs from the one in Eq.~\eqref{pad8a}.  
First of all, it contains the shift of the momentum ${\bm p}$ by $\frac{E_B^{\rm rel}-E_{\bm p}}{c}{\bm n}$, which is
responsible for the retardation effects~\cite{KKpress}. However, as it follows from our numerical results, it is justified to neglect this shift under current conditions. Moreover, 
it follows from Eq.~\eqref{Gnew} that the correction~\eqref{pd21} should be generalized such that $N({\bm p},{\bm k})=p\cdot k$,
which is accounted for in our modified QRSFA [i.e., its version ii) introduced in Sec.~\ref{sec:comparison}]. To have a close analogy between~\eqref{Gnew} and~\eqref{pad8a}, we also rewrite:
$E_{\bm p}-E_B^{\rm rel}=E_{\rm kin}({\bm p})-(E_B^{\rm rel}-m_{\rm e}c^2)$, with $E_{\rm kin}({\bm p})$ defined in~\eqref{pad11}. This suggest that when developing the modified QRSFA treatment 
one can account for the full relativistic kinetic energy of the electron~\eqref{pad11}. It also follows from here and the discussion
in Sec.~\ref{model} that, in the nonrelativistic limit, it is justified to replace $(E_B^{\rm rel}-m_{\rm e}c^2)$ by the binding energy of the hydrogen-like 
ion that is derived from the Schr\"odinger equation, $E_B$. Finally, while disregarding the retardation effects, we take $G(\omega t,{\bm p})$ in Eq.~\eqref{pad8}.

\end{document}